\documentclass[twocolumn,showpacs,preprintnumbers,showkeys,superscriptaddress]{revtex4}  %
\usepackage{amssymb}
\usepackage{graphicx}
\usepackage{dcolumn}
\usepackage{bm}
\usepackage{appendix}

\def\eqref#1{Eq.~(\ref{eq:#1})}

\begin{document}

\title{Solving for the Particle-Number-Projected HFB Wavefunction}
\author{L. Y. Jia}  \email{liyuan.jia@usst.edu.cn}
\affiliation{Department of Physics, University of Shanghai for
Science and Technology, Shanghai 200093, P. R. China}

\date{\today}

\begin{abstract}

Recently we proposed a particle-number-conserving theory for nuclear
pairing [Jia, Phys. Rev. C {\bf 88}, 044303 (2013)] through the
generalized density matrix formalism. The relevant equations were
solved for the case when each single-particle level has a distinct
set of quantum numbers and could only pair with its time-reversed
partner (BCS-type Hamiltonian). In this work we consider the more
general situation when several single-particle levels could have the
same set of quantum numbers and pairing among these levels is
allowed (HFB-type Hamiltonian). The pair condensate wavefunction
(the HFB wavefunction projected onto good particle number) is
determined by the equations of motion for density matrix operators
instead of the variation principle. The theory is tested in the
simple two-level model with factorizable pairing interactions and
the semi-realistic model with the zero-range delta interaction.

\end{abstract}

\pacs{ 21.60.Ev, 21.10.Re, }

\vspace{0.4in}

\maketitle

\section{Introduction}

The Bardeen-Cooper-Schrieffer (BCS) theory \cite{BCS} or its
advanced version the Hartree-Fock-Bogoliubov (HFB) theory
\cite{BCS_nucl2} has long been used to treat pairing correlations
\cite{BCS_nucl1} in atomic nuclei. In the mean field we introduce
the Bogoliubov quasi-particles and write the ground state as a
Slater determinant of the latter. Usually the variation principle is
used to determine the structure of the quasi-particles. Although
enjoying great success, the method has disadvantages of breaking the
exact particle number and a need for an unphysical minimum
pairing-force strength \cite{Bohr_book, Ring_book}.

The theory could be improved by the ``variation after
particle-number projection'' (VAPNP) procedure, in which the BCS or
HFB wavefunction is projected onto good particle number before the
variation is done \cite{Dietrich}. Effectively, the pair condensate
wavefunction [Eq. (\ref{gs})] is taken as the variational ground
state. The method has been discussed \cite{Dietrich, Flocard,
Sheikh_2000, Sheikh_2002, Robledo} and applied with success to
ultrasmall metallic grains \cite{Braun, Dukelsky, Delft} in the
VAPNP+BCS version, and to realistic nuclei \cite{Egido_1982,
Schmid_1987, Anguiano_2001, Anguiano_2002, Stoitsov_2007} in the
VAPNP+HFB version. However, the computing time cost is relatively
large and there is only a limited number of nuclei on the nuclear
chart that has been calculated by the VAPNP + HFB method in large
configuration spaces.

Recently we proposed \cite{Jia_2013_1} a new criteria to determine
the pair condensate wavefunction based on the Heisenberg equations
of motion for density matrix operators. The relevant equations have
been solved for the BCS-type Hamiltonian: Each single-particle level
has a distinct set of quantum numbers corresponding to the
symmetries of the pairing operator, thus a level could only be
paired with its time-reversed partner. The validity of the theory
has been proved on a large ensemble with random interactions
\cite{Jia_2013_2}. In this work we consider the general situation of
the HFB-type Hamiltonian, where several single-particle levels could
have the same set of quantum numbers and pairing among them is
allowed. The method should cost much less time in computing compared
with the traditional variational calculation. In the algorithm of
the current theory only the one-body density matrices on the pair
condensate are calculated, whereas the variation principle needs the
two-body density matrices in computing the expectation value of a
two-body Hamiltonian.

In Sec. \ref{sec_formalism} we give the general formalism to solve
for the particle-number-projected HFB wavefunction with the pairing
theory based on the generalized density matrix (GDM) formalism. Then
the method is tested in the simple two-level model with factorizable
pairing interactions and in the semi-realistic model with the
zero-range delta interaction, in Sec. \ref{sec_two_level} and Sec.
\ref{sec_delta_int}, respectively. Finally Sec. \ref{sec_summary}
summarizes the work.

\section{Formalism  \label{sec_formalism}}

The antisymmetrized (two-body) fermionic Hamiltonian governing the
dynamics of the system is written as
\begin{eqnarray}
H = \sum_{\alpha\beta} \epsilon_{\alpha\beta} a_\alpha^\dagger
a_\beta + \frac{1}{4} \sum_{\alpha\beta\gamma\delta}
V_{\alpha\beta\gamma\delta} a_\alpha^\dagger a_\beta^\dagger
a_\gamma a_\delta .  \label{H_f}
\end{eqnarray}
In the presence of pairing correlations, the ground state of the
$2N$-particle system is assumed to be an $N$-pair condensate,
\begin{eqnarray}
|\phi_N\rangle = \frac{1}{\sqrt{\chi_{N}}} (P^\dagger)^{N} |0\rangle
, \label{gs}
\end{eqnarray}
where $\chi_{N}$ is the normalization factor, $P^\dagger$ is the
pair creation operator
\begin{eqnarray}
P^\dagger = \sum_{\alpha\in\Lambda,\beta\in\Lambda} v_{\alpha\beta}
a_\alpha^\dagger a_{\tilde{\beta}}^\dagger . \label{P_dag_alpha}
\end{eqnarray}
In Eq. (\ref{P_dag_alpha}) $\Lambda$ represents a subspace
consisting of ``pair-indices'' whose dimension is half of that of
the single-particle space. We could take, for example, $\Lambda$ to
consist of those single-particle levels with a positive magnetic
quantum number. $|\tilde{\beta}\rangle$ is the time-reversed level
of the single-particle level $|\beta\rangle$
($|\tilde{\tilde{\beta}}\rangle = - |\beta\rangle$). The pair
structure matrix $v_{\alpha\beta}$ is Hermitian and block-diagonal.
The time-reversal invariance of $|\phi_N\rangle$ implies that
$v_{\beta\alpha} = v_{\alpha\beta}^*$ ($\hat{T} v_{\alpha\beta}
a_\alpha^\dagger a_{\tilde{\beta}}^\dagger |0\rangle =
v_{\alpha\beta}^* a_{\beta}^\dagger a_{\tilde{\alpha}}^\dagger
|0\rangle$). $v_{\alpha\beta}$ vanishes unless $|\alpha\rangle$ and
$|\beta\rangle$ have the same set $\tau$ of quantum numbers related
to the symmetries of $P^\dagger$. For example, in the spherical
shell model $\tau = (\pi, j, m)$ is a collection of parity $\pi$,
angular momentum $j$ and its projection $m$; in the deformed Nilsson
mean field $\tau = (\pi, K)$ is a collection of parity $\pi$ and
angular-momentum projection $K$ onto the symmetry axis.

We introduce the unitary transformation $\eta$ between the original
single-particle basis $\{|\alpha\rangle\}$ and the new
single-particle basis $\{|1\rangle\}$ as
\begin{eqnarray}
\eta_{\alpha1} \equiv \langle \alpha | 1 \rangle ,  \label{eta}
\end{eqnarray}
where elements of the transformation matrix $\eta$ are defined
through Dirac notation. Properties of time-reversal operation imply
that $(\eta^\dagger)_{1\alpha} = \langle 1 | \alpha \rangle =
\langle \tilde{\alpha} | \tilde{1} \rangle$. The matrix $\eta$ is
block-diagonal: $\eta_{\alpha1}$ vanishes unless $|1\rangle$ and
$|\alpha\rangle$ have the same quantum number $\tau$. Under the
transformation (\ref{eta}) the operator $P^\dagger$
(\ref{P_dag_alpha}) becomes
\begin{eqnarray}
P^\dagger = \sum_{1\in\Lambda,2\in\Lambda} v_{12} a_1^\dagger
a_{\tilde{2}}^\dagger , \label{P_dag_12}
\end{eqnarray}
with
\begin{eqnarray}
v_{12} = \sum_{\alpha\in\Lambda,\beta\in\Lambda}
(\eta^\dagger)_{1\alpha} v_{\alpha\beta} \eta_{\beta2} .
\label{v_12}
\end{eqnarray}
We choose the transformation $\eta$ that diagonalizes the Hermitian
matrix $v_{\alpha\beta}$ in Eq. (\ref{v_12}), consequently
$P^\dagger$ becomes
\begin{eqnarray}
P^\dagger = \sum_{1\in\Lambda} v_{1} a_1^\dagger
a_{\tilde{1}}^\dagger . \label{P_dag_1}
\end{eqnarray}
We call this new single-particle basis the ``canonical basis''. In
the following the Arabic numerals $|1\rangle$, $|2\rangle$, ...
refer to single-particle levels in this basis unless otherwise
specified.

In the canonical basis the density matrices for the pair condensate
(\ref{gs}) are ``diagonal'':
\begin{eqnarray}
\rho_{12} \equiv \langle \phi_{N} | a_2^\dagger a_1 | \phi_{N}
\rangle = \delta_{12} n_1 ,  \label{rho_def}  \\
\kappa_{12} \equiv \langle \phi_{N-1} | a_2 a_1 | \phi_{N} \rangle =
\delta_{\tilde{1}2} s_1 .  \label{kappa_def}
\end{eqnarray}
The normalization factor $\chi_N$ (\ref{gs}), occupation numbers
$n_1$ (\ref{rho_def}), and pair-transfer amplitudes $s_1$
(\ref{kappa_def}) are functions of the pair structures $v_1$
(\ref{P_dag_1}); their functional forms, as the ``kinematics'' of
the system, have already been given in Eqs. (23) and (24) of Ref.
\cite{Jia_2013_1} and are not repeated here. The Hartree-Fock mean
field $f$ and pairing mean field $\delta$ are defined as
\begin{eqnarray}
f_{12} \equiv \epsilon_{12} + \sum_{34} V_{1432} \rho_{34} =
\epsilon_{12} + \sum_{3} V_{1332} n_{3} ,  \label{f_def}  \\
\delta_{1\tilde{2}} \equiv \frac{1}{2} \sum_{34} V_{1\tilde{2}34}
\kappa_{43} = \frac{1}{2} \sum_{3} V_{1\tilde{2}\tilde{3}3} s_3 .
\label{delta_def}
\end{eqnarray}
$f$ and $\delta$ are block-diagonal matrix: $f_{12}$ and
$\delta_{1\tilde{2}}$ vanish unless $|1\rangle$ and $|2\rangle$ have
the same quantum number $\tau$. The Hamiltonian parameters
($\epsilon_{12}$ and $V_{1234}$) in Eqs. (\ref{f_def}) and
(\ref{delta_def}) should be calculated from those in the original
single-particle basis ($\epsilon_{\alpha\beta}$ and
$V_{\alpha\beta\gamma\delta}$) through the transformation
(\ref{eta}).

The equation of motion for the density matrix $\kappa$ has been
derived in Eq. (14) of Ref. \cite{Jia_2013_1},
\begin{eqnarray}
(E_N - E_{N-1}) \kappa = f \kappa + \kappa f^T + \delta - \delta
\rho^T - \rho \delta ,  \label{eom_K0}
\end{eqnarray}
where $E_N$ and $E_{N-1}$ are ground state energies for
$|\phi_N\rangle$ and $|\phi_{N-1}\rangle$, respectively. On the
right-hand side, $f$ and $\delta$ are the mean fields (\ref{f_def})
and (\ref{delta_def}), $f^T$ is the transpose of $f$, and terms like
`$f \kappa$' are understood as matrix multiplication. In deriving
Eq. (\ref{eom_K0}) we have used the main approximation of the
method,
\begin{eqnarray}
\langle \phi_{N-1} | a_4^\dagger a_3 a_2 a_1 | \phi_N \rangle
\nonumber \\
\approx \langle \phi_{N} | a_4^\dagger a_1 | \phi_{N}
\rangle
\langle \phi_{N-1} | a_3 a_2 | \phi_N \rangle  \nonumber \\
- \langle \phi_{N} | a_4^\dagger a_2 | \phi_{N} \rangle \langle
\phi_{N-1} | a_3 a_1 | \phi_N \rangle  \nonumber \\
+ \langle
\phi_{N} | a_4^\dagger a_3 | \phi_{N} \rangle \langle \phi_{N-1} |
a_2
a_1 | \phi_N \rangle  \nonumber \\
= \rho_{14} \kappa_{23} - \rho_{24} \kappa_{13} + \rho_{34}
\kappa_{12} ,  \label{fac2}
\end{eqnarray}
which says that on the pair condensate (\ref{gs}) the two-body
density matrix factorizes into products of one-body density matrices
in both the particle-hole and particle-particle channels.

Equation (\ref{eom_K0}) is a block-diagonal matrix equation; its
$1\tilde{2}$ matrix element vanishes unless $|1\rangle$ and
$|2\rangle$ have the same quantum number $\tau$. Within each block
we take the $1\tilde{2}$ matrix element on both sides of Eq.
(\ref{eom_K0}), when $1 \ne 2$ we get after simplification
\begin{eqnarray}
0 = (s_1 + s_2) f_{12} + (1 - n_1 - n_2) \delta_{1\tilde{2}} .
\label{main_12}
\end{eqnarray}
And the $1\tilde{1}$ matrix element gives
\begin{eqnarray}
E_N - E_{N-1} = 2 f_{11} + \delta_{1\tilde{1}} \frac{1 - 2 n_1}{s_1}
.  \label{main_11}
\end{eqnarray}
Equations (\ref{main_12}) and (\ref{main_11}) are the main equations
of the theory. Below we show that the number of constraints from
these two equations equals to the number of parameters in the pair
structure $v_{\alpha\beta}$ (\ref{P_dag_alpha}); thus the latter is
fixed completely.

We assume a single-particle space of dimension $2\Omega$ split by
the quantum number $\tau$ as $\Omega = \sum_\tau \Omega_\tau$. The
number of restrictions in Eq. (\ref{main_12}) is $\sum_\tau 2
C_{\Omega_\tau}^2 = \sum_\tau \Omega_\tau (\Omega_\tau - 1) =
\sum_\tau \Omega_\tau^2 - \Omega$ (there is a factor $2$ because the
equation has real and imaginary parts). Equation (\ref{main_11})
implies that the right-hand side is independent of the
single-particle label $1$, which gives $\Omega - 1$ constraints.
Hence the total number of constraints is $\sum_\tau \Omega_\tau^2 -
1$. On the other hand, the number of independent parameters in the
Hermitian block-diagonal pair-structure matrix $v_{\alpha\beta}$
(\ref{P_dag_alpha}) is $\sum_\tau \Omega_\tau^2 - 1$ (There is a
``$-1$'' because an overall normalization factor does not matter).
The number of restrictions indeed equals to the number of
parameters.

In practical calculations usually the Hamiltonian parameters
$\epsilon_{\alpha\beta}$ and $V_{\alpha\beta\gamma\delta}$
(\ref{H_f}) are real. In this case the pair structures
$v_{\alpha\beta}$ (\ref{P_dag_alpha}) could be taken as real numbers
and the transformation $\eta$ (\ref{eta}) is an orthogonal matrix.
The number of restrictions from Eqs. (\ref{main_12}) and
(\ref{main_11}) is $\sum_\tau C_{\Omega_\tau}^2 + \Omega - 1 =
\sum_\tau \frac{\Omega_\tau(\Omega_\tau+1)}{2} - 1$, which equals to
the number of independent parameters in the real symmetric
block-diagonal pair-structure matrix $v_{\alpha\beta}$
(\ref{P_dag_alpha}). In the following we assume that this is the
case.

An alternative parametrization may be more convenient in practice.
The independent parameters could be taken as those in the orthogonal
transformation $\eta$ (\ref{eta}) and the pair structure in the
canonical basis $v_{1}$ (\ref{P_dag_1}). The orthogonal
transformation for $\Omega_\tau = 2$ could be parameterized as
\begin{eqnarray}
\eta = \left(
  \begin{array}{cc}
    \eta_{\alpha 1} & \eta_{\alpha 2} \\
    \eta_{\beta 1} & \eta_{\beta 2} \\
  \end{array}
\right) = \left(
  \begin{array}{cc}
    \cos \theta   & - \sin \theta \\
    \sin \theta & \cos \theta \\
  \end{array}
\right) .  \label{tran2}
\end{eqnarray}
For $\Omega_\tau = 3$,
\begin{eqnarray}
\eta = \eta^{(12)} \eta^{(13)} \eta^{(23)} = \left(
  \begin{array}{ccc}
    \cos\theta_{12} & -\sin\theta_{12} & 0  \\
    \sin\theta_{12} & \cos\theta_{12} & 0  \\
    0 & 0 & 1  \\
  \end{array}
\right)  \nonumber \\
\cdot \left(
  \begin{array}{ccc}
    \cos\theta_{13} & 0 & -\sin\theta_{13} \\
    0 & 1 & 0  \\
    \sin\theta_{13} & 0 & \cos\theta_{13} \\
  \end{array}
\right) \left(
  \begin{array}{ccc}
    1 & 0 & 0 \\
    0 & \cos\theta_{23} & -\sin\theta_{23} \\
    0 & \sin\theta_{23} & \cos\theta_{23} \\
  \end{array}
\right) .     \nonumber
\end{eqnarray}
And in general,
\begin{eqnarray}
\eta = \prod_{i<j} \eta^{(ij)} .  \label{eta_para}
\end{eqnarray}

\section{Two-Level Model  \label{sec_two_level}}

We test the formalism in the simple model with two single-particle
levels and a factorizable pairing interaction. The theory is solved
analytically and the results are compared with the exact shell-model
results.

We assume the rotational invariance and a single-particle space of
two $j$-levels each with degeneracy $2\Omega = 2j+1$ (degenerate in
the magnetic quantum number $m$). The fermionic Hamiltonian
(\ref{H_f}) takes the form
\begin{eqnarray}
H = \sum_m ( \epsilon_\alpha a_{\alpha m}^\dagger a_{\alpha m} +
\epsilon_\beta a_{\beta m}^\dagger a_{\beta m} ) - \Pi^\dagger \Pi
 , \label{H_alphabeta}
\end{eqnarray}
where $\Pi^\dagger$ is the pairing operator
\begin{widetext}
\begin{eqnarray}
\Pi^\dagger = \sum_{m>0} ( g a_{\alpha m}^\dagger a_{\alpha
\tilde{m}}^\dagger + g a_{\beta m}^\dagger a_{\beta
\tilde{m}}^\dagger + p a_{\alpha m}^\dagger a_{\beta
\tilde{m}}^\dagger + p a_{\beta m}^\dagger a_{\alpha
\tilde{m}}^\dagger ) ,  \label{Pi_alphabeta}
\end{eqnarray}
in which $g$ and $p$ are ``diagonal'' and ``off-diagonal'' pairing
strengths, respectively. The Hamiltonian parameters in the canonical
single-particle basis are calculated through the transformation
(\ref{tran2}),
\begin{eqnarray}
H = \sum_m \sum_{ij} \epsilon_{ij} a_{i m}^\dagger a_{j m} -
\Pi^\dagger \Pi
 , \label{H_12}
\end{eqnarray}
where
\begin{eqnarray}
\left(
  \begin{array}{cc}
    \epsilon_{11} & \epsilon_{12} \\
    \epsilon_{21} & \epsilon_{22} \\
  \end{array}
\right) = \left(
  \begin{array}{cc}
    \epsilon_\alpha \cos^2\theta + \epsilon_\beta \sin^2\theta & \frac{1}{2} (\epsilon_\beta-\epsilon_\alpha) \sin2\theta  \\
    \frac{1}{2} (\epsilon_\beta-\epsilon_\alpha) \sin2\theta & \epsilon_\alpha \sin^2\theta + \epsilon_\beta \cos^2\theta  \\
  \end{array}
\right) ,  \label{epsilon_12}
\end{eqnarray}
and
\begin{eqnarray}
\Pi^\dagger = \sum_{m>0} [ (g + p \sin2\theta) a_{1 m}^\dagger a_{1
\tilde{m}}^\dagger + (g - p \sin2\theta) a_{2 m}^\dagger a_{2
\tilde{m}}^\dagger + p \cos2\theta a_{1 m}^\dagger a_{2
\tilde{m}}^\dagger + p \cos2\theta a_{2 m}^\dagger a_{1
\tilde{m}}^\dagger ] . \label{Pi_12}
\end{eqnarray}
Based on Eqs. (\ref{f_def}) and (\ref{delta_def}) we compute the
matrix elements of the mean fields in the canonical single-particle
basis,
\begin{eqnarray}
f_{11} = \epsilon_\alpha \cos^2\theta + \epsilon_\beta \sin^2\theta
- p^2 \cos^2 2\theta n_2 - (g + p \sin2\theta)^2 n_1 ,
  \nonumber \\
f_{22} = \epsilon_\alpha \sin^2\theta + \epsilon_\beta \cos^2\theta
- ( g - p \sin2\theta )^2 n_2 - p^2 \cos^2 2\theta n_1 ,
\nonumber \\
f_{12} = \frac{1}{2}(\epsilon_\beta-\epsilon_\alpha) \sin2\theta -
p\cos2\theta [ (g - p \sin2\theta) n_2 + (g + p \sin2\theta) n_1 ] ,
\nonumber
\end{eqnarray}
and
\begin{eqnarray}
\delta_{1\tilde{1}} = - \Omega (g + p \sin2\theta) [(g + p
\sin2\theta) s_1 + (g - p \sin2\theta) s_2 ] ,  \nonumber \\
\delta_{2\tilde{2}} = - \Omega (g - p \sin2\theta) [(g + p
\sin2\theta) s_1 + (g - p \sin2\theta) s_2 ] ,  \nonumber \\
\delta_{1\tilde{2}} = - \Omega p \cos2\theta [ (g + p \sin2\theta)
s_1 + (g - p \sin2\theta) s_2 ] .     \nonumber
\end{eqnarray}
Consequently Eqs. (\ref{main_12}) and (\ref{main_11}) imply
\begin{eqnarray}
0 = \sin2\theta (s_1 + s_2) (\epsilon_\beta-\epsilon_\alpha) + 2 g p
\cos2\theta (s_1+s_2) [(n_1 + n_2) (\Omega-1) -
\Omega ]  \nonumber \\
+ p^2 \sin4\theta [ \Omega (n_1 + n_2 - 1) (s_1 - s_2) - (s_1 + s_2)
(n_1 - n_2) ] ,  \label{main_2l_12}
\end{eqnarray}
and
\begin{eqnarray}
0 = \cos2\theta s_1 s_2 (\epsilon_\alpha - \epsilon_\beta) + p^2
\cos^2 2\theta s_1 s_2 (n_1 - n_2) + \frac{1}{2} \Omega (g^2 - p^2
\sin^2 2\theta) [ (s_1)^2 (1 - 2 n_2) - (s_2)^2
(1 - 2 n_1) ]  \nonumber \\
- 2 g p \Omega \sin2\theta s_1 s_2 + ( \Omega - 1 ) (g + p
\sin2\theta)^2 n_1 s_1 s_2 - ( \Omega - 1 ) ( g - p \sin2\theta )^2
n_2 s_1 s_2 . \label{main_2l_11}
\end{eqnarray}
These two equations fix the two independent model parameters
$v_2/v_1$ and $\theta$ ($n_1$ and $s_1$ are functions of $v_2/v_1$).
Then the density matrices in the original single-particle basis are
calculated through the transformation $\eta$ (\ref{tran2}),
\begin{eqnarray}
\left(
  \begin{array}{cc}
    \rho_{\alpha\alpha} & \rho_{\alpha\beta} \\
    \rho_{\beta\alpha} & \rho_{\beta\beta} \\
  \end{array}
\right) = \left(
  \begin{array}{cc}
    n_1 \cos^2\theta + n_2 \sin^2\theta & (n_1-n_2) \sin\theta \cos\theta \\
    (n_1-n_2) \sin\theta \cos\theta & n_1 \sin^2\theta + n_2 \cos^2\theta \\
  \end{array}
\right) ,  \label{rho_ori}
\end{eqnarray}
and
\begin{eqnarray}
\left(
  \begin{array}{cc}
    \kappa_{\alpha\tilde{\alpha}} & \kappa_{\alpha\tilde{\beta}} \\
    \kappa_{\beta\tilde{\alpha}} & \kappa_{\beta\tilde{\beta}} \\
  \end{array}
\right) = \left(
  \begin{array}{cc}
    s_1 \cos^2\theta + s_2 \sin^2\theta & (s_1-s_2) \sin\theta \cos\theta \\
    (s_1-s_2) \sin\theta \cos\theta & s_1 \sin^2\theta + s_2 \cos^2\theta \\
  \end{array}
\right) .  \label{kappa_ori}
\end{eqnarray}
\end{widetext}

We consider the range of model parameters in the realistic spherical
nuclear shell model. Usually the two single-particle levels
$|\alpha\rangle$ and $|\beta\rangle$ belong to different major
shells and $\epsilon_\beta - \epsilon_\alpha$ is bigger than $10$
MeV in magnitude that is about the energy of two major-shell gaps
(adjacent major shells have opposite parity). Based on the empirical
pairing strength formula $G = g^2 \approx 20/A$ MeV ($A$ is the mass
number) \cite{Nilsson_1961}, $G = g^2$ should be less than $2$ MeV
in medium and heavy nuclei having considered possible deviations
from the constant pairing formula. The off-diagonal pairing strength
$p$ is usually smaller than the diagonal pairing strength $g$ owing
to the smaller overlap of the two single-particle wavefunctions
involved.

We test the model numerically in an ensemble consisting of examples
with different model parameters. The single-particle angular
momentum $j$ of the examples takes value from $j = \frac{3}{2},
\frac{5}{2}, \frac{7}{2}, \frac{9}{2}$. The number of pairs is in
the range $1 \le N \le 2j = 2\Omega-1$ (from two particles to two
holes in the model space). The energy gap between the two
single-particle levels is selected to be the energy unit so that
$\epsilon_\alpha = -0.5$ and $\epsilon_\beta = 0.5$. Based on the
estimations in the previous paragraph, we choose the range of the
pairing strength to be $0 < g \le 0.5$ and $0 < p \le g$ with step
size $0.1$. We scan the whole parameter space and the total number
of examples in the ensemble is $15 \sum_j 2j = 15 \times 24 = 360$
(the factor of $15$ is the number of possible values of $g$ and
$p$).

\begin{figure}
\includegraphics[width = 0.5\textwidth]{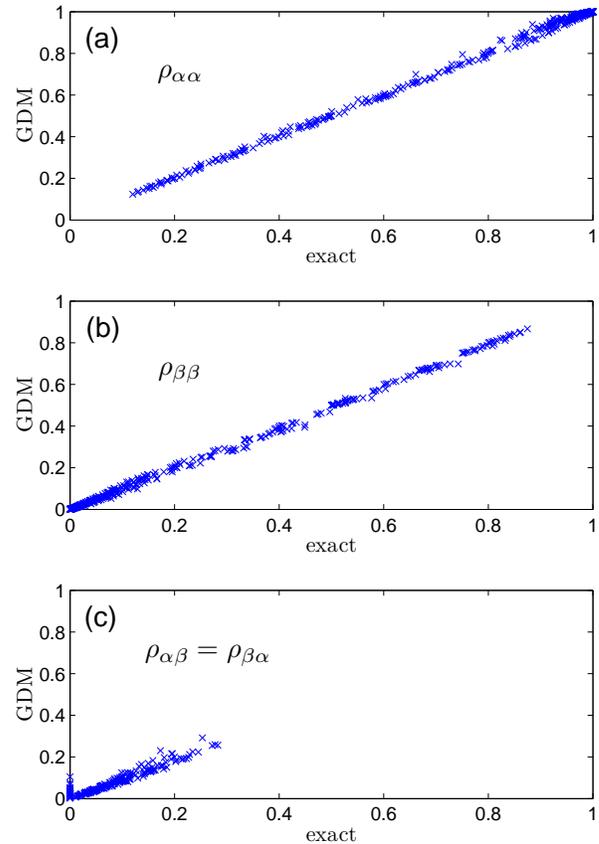}
\caption{\label{Fig_rho_line} (Color online) Density matrix $\rho$
in the original single-particle basis for the ensemble. In panel (a)
there are $360$ points corresponding to the $360$ examples in the
ensemble. The horizontal coordinate of each point is
$\rho_{\alpha\alpha}$ of the example by the exact diagonalization
and the vertical coordinate is the one by the GDM calculation.
Similarly for panels (b) and (c), but the coordinates of the points
are $\rho_{\beta\beta}$ and $\rho_{\alpha\beta}$ of the examples,
respectively. }
\end{figure}

\begin{figure}
\includegraphics[width = 0.5\textwidth]{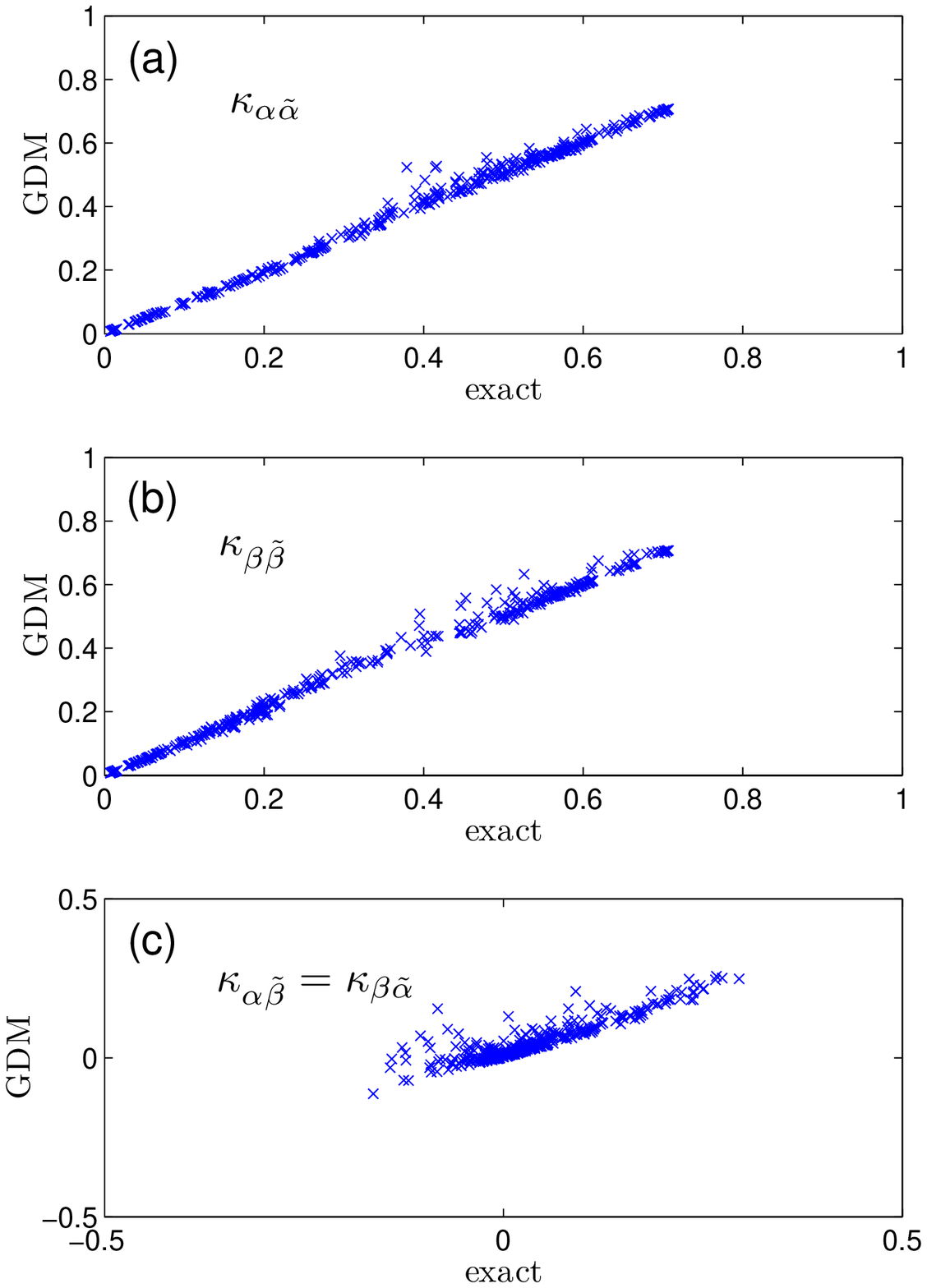}
\caption{\label{Fig_kappa_line} (Color online) Density matrix
$\kappa$ in the original single-particle basis for the ensemble. In
panel (a) there are $360$ points corresponding to the $360$ examples
in the ensemble. The horizontal coordinate of each point is
$\kappa_{\alpha\tilde{\alpha}}$ of the example by the exact
diagonalization and the vertical coordinate is the one by the GDM
calculation. Similarly for panels (b) and (c), but the coordinates
of the points are $\kappa_{\beta\tilde{\beta}}$ and
$\kappa_{\alpha\tilde{\beta}}$ of the examples, respectively. }
\end{figure}

\begin{figure}
\includegraphics[width = 0.5\textwidth]{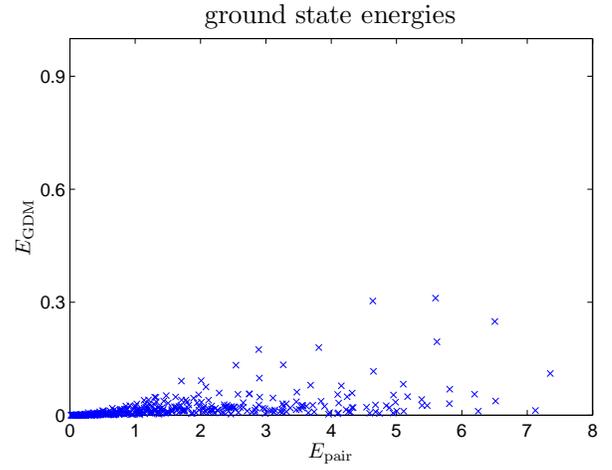}
\caption{\label{Fig_E_line} (Color online) Ground state energies in
the ensemble. In the figure there are $360$ points corresponding to
the $360$ examples in the ensemble. The horizontal coordinate of
each point is the pairing correlation energy $E_{\rm{pair}}$ of the
example. The vertical coordinate $E_{\rm{GDM}}$ is the ground state
energy by the GDM calculation, measured from the exact ground state
energy. }
\end{figure}

The results for the density matrices $\rho$ and $\kappa$ in the
original single-particle basis are shown in Figs. \ref{Fig_rho_line}
and \ref{Fig_kappa_line}, respectively. Each point in these figures
corresponds to one example in the ensemble with the horizontal
coordinate being the exact result of the respective quantity and the
vertical one being that from the GDM calculation, thus a perfect
calculation would have all the points lying on the $y=x$ straight
line. From Figs. \ref{Fig_rho_line} and \ref{Fig_kappa_line} we see
that in general the GDM theory reproduces the exact density matrices
well. The root-mean-square deviations are
\begin{eqnarray}
\sigma_{\rho_{\alpha\alpha}} = \sigma_{\rho_{\beta\beta}} = 0.0125
,~ \sigma_{\rho_{\alpha\beta}} = \sigma_{\rho_{\beta\alpha}} =
0.0198 ,  \nonumber
\\
\sigma_{\kappa_{\alpha\tilde{\alpha}}} = 0.0183 ,~
\sigma_{\kappa_{\beta\tilde{\beta}}} = 0.0214 ,~
\sigma_{\kappa_{\alpha\tilde{\beta}}} =
\sigma_{\kappa_{\beta\tilde{\alpha}}} = 0.0384 .  \nonumber
\end{eqnarray}
$\sigma_{\rho_{\alpha\alpha}}$ and $\sigma_{\rho_{\beta\beta}}$ are
equal because $\rho_{\alpha\alpha} + \rho_{\beta\beta} = N/\Omega$.
The results for the ground state energies are shown in Fig.
\ref{Fig_E_line}. The horizontal coordinate of each point is the
pairing correlation energy for the example $E_{\rm{pair}} \equiv
\sum_{1} \epsilon_1 n^F_1 - E_{\rm{exact}}$, where $E_{\rm{exact}}$
is the exact ground state energy by the direct diagonalization, and
$n^F_1 = 1$ or $0$ is the occupation number of the naive Fermi
distribution. The vertical coordinate is the ground state energy by
the GDM calculation measured from the exact one, $E_{\rm{GDM}} =
\langle \phi_N | H | \phi_N \rangle - E_{\rm{exact}}$, where
$\langle \phi_N | H | \phi_N \rangle$ is calculated in the canonical
single-particle basis using the Hamiltonian (\ref{H_12}) together
with the recursive formulas derived in Refs.
\cite{Jia_2013_1,Jia_arXiv_2014}. We see that in general the GDM
calculation reproduces well the ground state energies: the errors
are small compared with the pairing correlation energies. The
average values of the errors and the pairing correlation energies
for the ensemble are $\bar{E}_{\rm{GDM}} = 0.0198$ and
$\bar{E}_{\rm{pair}} = 1.78$, respectively.

\begin{figure}
\includegraphics[width = 0.5\textwidth]{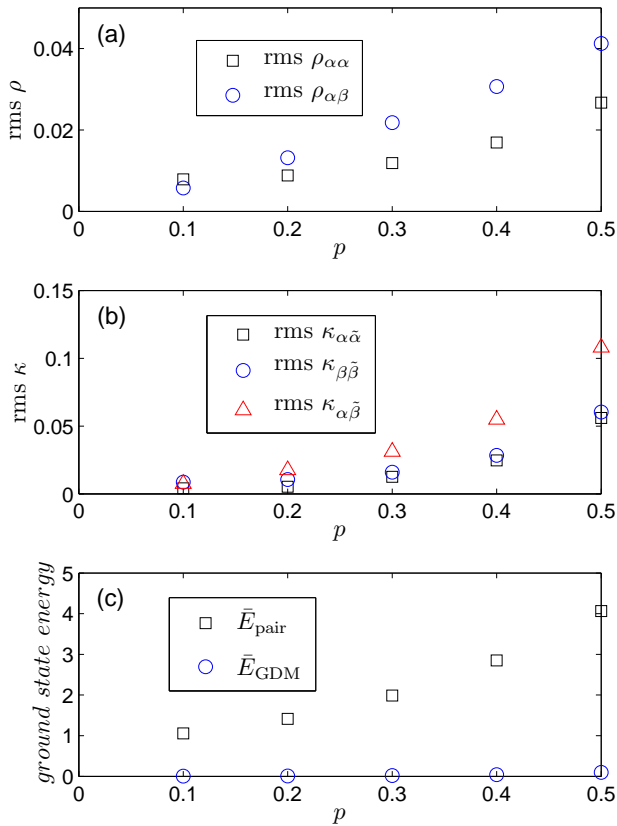}
\caption{\label{Fig_derivations_p} (Color online) Deviations of
various quantities from the exact results as functions of the
off-diagonal pairing strength $p$. The $360$ examples in the
ensemble are divided into five subgroups according to the value of
$p$. Panel (a) plots the root-mean-square deviations from the exact
results of the density matrix $\rho$ by the GDM calculation within
each subgroup, where black squares and blue circles are for
$\rho_{\alpha\alpha}$ and $\rho_{\alpha\beta}$, respectively.
Similarly panel (b) plots the root-mean-square deviations of the
density matrix $\kappa$ within each subgroup, where black squares,
blue circles, and red triangles are for
$\kappa_{\alpha\tilde{\alpha}}$, $\kappa_{\beta\tilde{\beta}}$, and
$\kappa_{\alpha\tilde{\beta}}$, respectively. In panel (c) the black
squares show the average value of the pairing correlation energy
$\bar{E}_{\rm{pair}}$ within each subgroup, and the blue circles
show the average value of the ground state energy by the GDM
calculations $\bar{E}_{\rm{GDM}}$, measured from the exact one. }
\end{figure}

At last in Fig. \ref{Fig_derivations_p} we show the accuracy of the
method depending on the off-diagonal pairing strength $p$. The $360$
examples in the ensemble are divided into five subgroups according
to the value of $p$, and deviations within each subgroup of various
quantities are calculated. We see that in general the errors of the
GDM calculation increase with $p$. In realistic spherical nuclear
shell model $p$ should be smaller than about $0.3$ based on the
estimations made in the paragraph below Eq. (\ref{kappa_ori}). In
this region the GDM theory is rather accurate.

\section{Model with Zero-Range Delta Interaction  \label{sec_delta_int}}

In this section we further test the theory in a semi-realistic model
with the pairing Hamiltonian matrix elements calculated from the
zero-range delta interaction. The model has only one species of
nucleons (for example, neutrons) and the single-particle space
consists of five levels: three of them ($2s_{1/2}$, $1d_{3/2}$,
$1d_{5/2}$) are from the $4 \hbar \omega$ major shell and two
($3s_{1/2}$, $2d_{5/2}$) are from the $6 \hbar \omega$ major shell.
The single-particle energies are determined by random number
generator to be $\epsilon_{2s_{1/2}} = 1.16$, $\epsilon_{1d_{3/2}} =
2.11$, $\epsilon_{1d_{5/2}} = 0.11$, $\epsilon_{3s_{1/2}} = 12.98$,
and $\epsilon_{2d_{5/2}} = 10.37$ MeV (The gap between the $4 \hbar
\omega$ and $6 \hbar \omega$ major shells is taken to be around $10$
MeV). The two-body pairing matrix elements [$\langle \alpha
\beta;J|V|\gamma \delta;J\rangle$ with $J=0$] are calculated from
the zero-range delta interaction $V(\vec{r}_1,\vec{r}_2) = - \lambda
\delta(\vec{r}_1 - \vec{r}_2)$, taking the single-particle
wavefunctions ($|\alpha\rangle$, $|\beta\rangle$ ...) to be the
harmonic oscillator ones. We perform six sets of calculations at
different model parameters of the particle number $2N$ and the
pairing strength $\lambda$ as listed in table \ref{tab_setpara}. The
unit for $\lambda$ is ${\rm{MeV}} \cdot b^3$, where $b =
\sqrt{\hbar/(\mu \omega)}$ is the length parameter for the harmonic
oscillator wavefunctions ($\mu$ is the mass of the nucleon). Values
of $\lambda$ are chosen so that magnitudes of the resulting two-body
pairing matrix elements are realistic. The last row of table
\ref{tab_setpara} shows the average value of the pairing matrix
elements of a specific form,
${\rm{mean}}[-V_{\alpha\tilde{\alpha}\tilde{\alpha}\alpha}] =
-\sum_\alpha V_{\alpha\tilde{\alpha}\tilde{\alpha}\alpha} /
(2\Omega)$, where $\alpha$ runs over the entire single-particle
space whose dimension is $2\Omega$.

The results are shown in table \ref{tab_res}. We see that in general
the GDM calculation reproduces the exact density matrices
accurately, except for the very small value of $\rho_{\beta\beta}$
in set5. Particularly, the abrupt change of $\kappa$ values from
set3 to set4 is captured very well. The off-diagonal parts of the
density matrices ($\rho_{\alpha\mu}$, $\rho_{\gamma\nu}$,
$\kappa_{\alpha\tilde{\mu}}$, and $\kappa_{\gamma\tilde{\nu}}$),
originating from the ``off-diagonal'' parts of the HFB-type pairing
Hamiltonian, are always well reproduced, ranging from a
less-than-one-percent effect to a few. The errors of the GDM ground
state energy relative to the exact ones are shown in the last
column. We see that the errors are small and consistent with the
statistical estimate made within a large random ensemble for the
case of a BCS-type pairing Hamiltonian in Ref. \cite{Jia_2013_2}.

\section{Summary  \label{sec_summary}}

In summary, we considered the solution of the GDM pairing theory
with the HFB-type Hamiltonian. In the algorithm only the one-body
density matrix is computed, thus the method should be much faster
than the VAPNP + HFB method in which the two-body density matrix is
needed in calculating the expectation value of a two-body
Hamiltonian. With the assumption that the parent and daughter nuclei
were represented by pair condensates with the same pair structure,
the pair-transfer amplitudes $\kappa$ could be calculated in one run
together with the occupation numbers $\rho$. In contrast, the
traditional variation method calculates the parent and daughter
nuclei separately, each with a different pair structure.

The formalism is tested in the simple two-level model with
factorizable pairing interactions, and the semi-realistic model with
the zero-range delta interaction. In both cases the GDM calculation
reproduces quite well the exact density matrices and ground state
energy within the physical range of parameters for the realistic
spherical nuclear shell model. The errors are small and consistent
with the statistical estimates made within a large random ensemble
for the BCS-type Hamiltonian in Ref. \cite{Jia_2013_2}. In the
current form the theory is ready for the application to realistic
nuclear systems.

\section{Acknowledgments}

Support is acknowledged from the Hujiang Foundation of China
(B14004), and the startup funding for new faculty member in
University of Shanghai for Science and Technology. Part of the
calculations is done at the High Performance Computing Center of
Michigan State University.

\newpage

\newpage
\phantom{a}
\newpage

\newpage
\phantom{a}
\newpage

\newpage
\phantom{a}
\newpage

\newpage
\phantom{a}
\newpage

\begin{table}
  \centering
  \caption{Particle number $2N$ and pairing strength $\lambda$ as
  input parameters for the six sets of calculations.
  The quantity $b = \sqrt{\hbar/(\mu \omega)}$ in the unit of
  $\lambda$ is the length parameter for the harmonic oscillator wavefunctions.
  Also shown is ${\rm{mean}}[-{V}_{\alpha\tilde{\alpha}\tilde{\alpha}\alpha}]$, the average value of the pairing matrix
  elements of a specific form.
  See text for details.}
  \label{tab_setpara}
  \begin{tabular}{|c|c|c|c|c|c|c|}
  \hline
  ~         & set1 & set2 & set3 & set4 & set5 & set6  \\
  \hline
  $N$       & 2   & 3   & 4     & 5     & 4     & 4 \\ \hline
  $\lambda /({\rm{MeV}} \cdot b^3)$ & 20 & 20     & 20     & 20     & 10     &
  40  \nonumber  \\ \hline
  ${\rm{mean}}[-{V}_{\alpha\tilde{\alpha}\tilde{\alpha}\alpha}] / {\rm{MeV}}$ & 0.356 & 0.356     & 0.356     & 0.356     & 0.178     & 0.711 \\
  \hline
\end{tabular}
\end{table}

%

\newpage
\phantom{a}
\newpage

\begin{table*}
  \centering
  \caption{Density matrices $\rho$ and $\kappa$ by the exact and GDM
  calculations in the original single-particle basis. For clarity each single-particle level
  is represented by a Greek letter as $\{\alpha,\beta,\gamma,\mu,\nu\} =
  \{2s_{1/2}, 1d_{3/2}, 1d_{5/2}, 3s_{1/2}, 2d_{5/2}\}$. For each
  set of calculation the results of density matrices $\rho$ or $\kappa$
  are listed in two rows, with the top row being the exact results and
  the bottom row the GDM results. The last column gives the errors
  of the GDM ground state energy relative to the exact one, $E_{\rm{GDM}} =
\langle \phi_N | H | \phi_N \rangle - E_{\rm{exact}}$.}
  \label{tab_res}
\begin{tabular}{|c||c|c|c|c|c|c|c||c|c|c|c|c|c|c||c|}
  \hline
  ~     & $\rho_{\alpha\alpha}$ & $\rho_{\beta\beta}$ & $\rho_{\gamma\gamma}$ & $\rho_{\mu\mu}$ & $\rho_{\nu\nu}$ & $\rho_{\alpha\mu}$ & $\rho_{\gamma\nu}$ & $\kappa_{\alpha\tilde{\alpha}}$ & $\kappa_{\beta\tilde{\beta}}$ & $\kappa_{\gamma\tilde{\gamma}}$ & $\kappa_{\mu\tilde{\mu}}$ & $\kappa_{\nu\tilde{\nu}}$ & $\kappa_{\alpha\tilde{\mu}}$ & $\kappa_{\gamma\tilde{\nu}}$ & $E_{\rm{GDM}}/{\rm{keV}}$
  \\ \hline
set1 & 0.0400 & 0.0278 & 0.6326 & 0.0007 & 0.0020 & 0.0046 & 0.0285 & 0.1972 & 0.1655 & 0.6554 & 0.0162 & 0.0275 & 0.0213 & 0.0282 & ~ \\
   ~ & 0.0536 & 0.0313 & 0.6261 & 0.0008 & 0.0016 & 0.0056 & 0.0227 & 0.2283 & 0.1760 & 0.6524 & 0.0171 & 0.0282 & 0.0224 & 0.0227 & 7.44 \\ \hline
set2 & 0.0889 & 0.0342 & 0.9460 & 0.0009 & 0.0013 & 0.0078 & 0.0233 & 0.2910 & 0.1819 & 0.5837 & 0.0178 & 0.0265 & 0.0243 & 0.0136 & ~ \\
   ~ & 0.1086 & 0.0373 & 0.9374 & 0.0010 & 0.0011 & 0.0090 & 0.0200 & 0.3217 & 0.1919 & 0.5778 & 0.0186 & 0.0271 & 0.0253 & 0.0117 & 9.95 \\ \hline
set3 & 0.9766 & 0.0420 & 0.9776 & 0.0030 & 0.0012 & 0.0479 & 0.0216 & 0.9418 & 0.1482 & 0.1932 & 0.0276 & 0.0163 & 0.0449 & 0.0039 & ~ \\
   ~ & 0.9423 & 0.0312 & 0.9968 & 0.0029 & 0.0007 & 0.0454 & 0.0183 & 0.9041 & 0.1762 & 0.2148 & 0.0283 & 0.0178 & 0.0423 & 0.0036 & 129 \\ \hline
set4 & 0.9867 & 0.5254 & 0.9851 & 0.0030 & 0.0014 & 0.0478 & 0.0204 & 0.1419 & 0.7048 & 0.1455 & 0.0141 & 0.0249 & 0.0062 & 0.0024 & ~ \\
   ~ & 0.9934 & 0.5085 & 0.9947 & 0.0025 & 0.0010 & 0.0478 & 0.0189 & 0.1277 & 0.7000 & 0.1367 & 0.0138 & 0.0248 & 0.0055 & 0.0021 & 62.9 \\
   \hline\hline
set5 & 0.9952 & 0.0078 & 0.9959 & 0.0007 & 0.0002 & 0.0231 & 0.0096 & 0.9930 & 0.0519 & 0.0576 & 0.0124 & 0.0046 & 0.0227 & 0.0005 & ~ \\
   ~ & 0.9935 & 0.0029 & 0.9999 & 0.0007 & 0.0001 & 0.0228 & 0.0089 & 0.9914 & 0.0540 & 0.0586 & 0.0127 & 0.0046 & 0.0225 & 0.0005 & 42.1 \\ \hline
set6 & 0.9032 & 0.2134 & 0.8775 & 0.0139 & 0.0079 & 0.0990 & 0.0541 & 0.6378 & 0.4148 & 0.4880 & 0.0588 & 0.0652 & 0.0641 & 0.0255 & ~ \\
   ~ & 0.8407 & 0.2428 & 0.8795 & 0.0131 & 0.0074 & 0.0931 & 0.0471 & 0.5500 & 0.4623 & 0.5086 & 0.0587 & 0.0708 & 0.0553 & 0.0236 & 76.7 \\ \hline
\end{tabular}
\end{table*}

\end{document}